# Quantum Tsallis entropy and projective measurement

Marko V. Jankovic

*Tokyo Institute of Technology, 2-12-1 Ookayama, Meguro-ku, Tokyo, 152-8552, Japan*

It is well known that projective measurement will not decrease the von Neumann entropy of a quantum state. In this paper, it is shown that projective measurement will not decrease the quantum Tsallis entropy of a quantum state, either. Using a similar analysis, it can be shown that projective measurement will not decrease the quantum unified (*r*, *s*)-entropy in general.

## I. INTRODUCTION

The nonadditive quantum information theory is an attempt of generalizing the ordinary quantum information theory by introducing the quantum-mechanical version of the Tsallis entropy[1] and its related quantities. The mathematical foundations of the classical Tsallis entropy can be found in (Refs. 2 and 3). The Tsallis entropy represents entropy measure which is suitable for non-extensive systems as it is the case of the evolution of star clusters or in systems with long range microscopic memory, in fractal- or multifractal-like and self-organized critical systems, etc. The nonadditive quantum information theory was recently applied on several problems. An example is the problem of measuring the degree of state purification[4]. Another implementation is in the area of artificial neural networks[5]. In the neural networks area, a generation of useful algorithms for principal component analysis, minor component analysis and independent component analysis can be based on optimization of the quantum Tsallis entropy. Some existing algorithms can be seen in the light of minimization of the quantum Tsallis entropy of the state obtained from the mixture state after projection measurement.

In this paper, we analyze the behavior of the Tsallis entropy of a quantum system when we perform projective measurement on that system. In Sec. II, we present the definitions of quantum entropies. The main result of this paper is presented in Sec. III. In that section it is proved that the quantum Tsallis entropy will not be decreased by projective measurement. This result can be extended to the quantum (*r*, *s*)-entropy[6] in general.

## II. QUANTUM ENTROPY

This section is based on a similar section in (Ref. 6). Von Neumann defined the entropy of a quantum state $\rho$ by the following formula

$$S(\rho) = -\text{tr}(\rho \log(\rho)). \tag{1}$$

The quantum probability model takes place in a Hilbert space H of finite or infinite dimension. A state is represented by a positive semidefinite linear mapping (a matrix $\rho$) from this space to itself, with a trace of 1, i.e. $\forall \Psi \in H, \Psi^T \rho \Psi \geq 0$, tr($\rho$) =1. Such $\rho$ is selfadjoint and is called a density matrix. There are also some other, parameterized, definitions of quantum entropy that could be considered as generalizations of the von Neumann entropy. We will list some of them here:
Quantum Renyi entropy of order *r*,

$$S_r(\rho) = (1-r)^{-1} \log(\text{tr}(\rho^r)), \quad r > 0, r \neq 1. \tag{2}$$

Quantum entropy of degree *r* (quantum Tsallis entropy of degree *q*=*r*),

$$S_q(\rho) = S^r(\rho) = (1-r)^{-1}(\text{tr}(\rho^r) - 1), \quad r > 0, r \neq 1. \tag{3}$$

Quantum entropy of type *r*,

$$_r S(\rho) = (r-1)^{-1}\left(\left[\text{tr}\left(\rho^{1/r}\right)\right]^r - 1\right), \quad r > 0, r \neq 1. \tag{4}$$

Quantum (*r*, *s*) entropy

$$S_r^s(\rho) = [(1-r)s]^{-1}\left(\left(\mathrm{tr}(\rho^r)\right)^s - 1\right), \quad r > 0, r \neq 1. \tag{5}$$

Quantum unified $(r, s)$-entropy,

$$E_r^s = \begin{cases} S_r^s(\rho) & \text{if } r \neq 1, s \neq 0 \\ S_r(\rho) & \text{if } r \neq 1, s = 0 \\ S^r(\rho) & \text{if } r \neq 1, s = 1 \\ {}_rS(\rho) & \text{if } r \neq 1, s = 1/r \\ S(\rho) & \text{if } r = 1, \end{cases} \tag{6}$$

for all $r > 0$, and for any $s$.

It can be shown that the following equation holds

$$\lim_{r \to 1} S_r(\rho) = \lim_{r \to 1} S^r(\rho) = \lim_{r \to 1} {}_rS(\rho) = S(\rho).$$

In the following text, we will mainly address the Tsallis entropy, with the adopted notation $S_q(\rho)$.

### III. TSALLIS ENTROPY AND PROJECTIVE MEASUREMENT

How does the Tsallis entropy of a quantum system behave when we perform measurement on that system? Not surprisingly, like it is the case with the von Neumann entropy, the answer to this question depends on the type of measurement which we perform. However, we can have some general assertions, like it is the case with the von Neumann entropy. It is known that in the case of von Neumann entropy, projective measurement cannot decrease the entropy of the quantum system on which it is applied[7].

Here, we will examine what is going to happen if we perform projective measurement on the system. Such a measurement is a special kind of positive trace-preserving quantum operation. Consider an observable, $Q$, with eigenspaces defined by orthogonal projections, $\{P_k\}$, and measured values $\{q_k\}$, $Q$ is expressed as

$$Q = \sum_k q_k P_k, \tag{7}$$

Where $P_k P_{k'} = \delta_{kk'}$ and $\sum_k P_k = I$, with the identity operator $I$. The probability of finding the value $q_k$ of $Q$ in a state $\rho$ via the projective measurement reads $p_k = \mathrm{tr}(\rho P_k)$. $\rho$ is mapped to $\rho_k = (p_k)^{-1} P_k \rho P_k$. Then averaging over all possible outcomes, we obtain

$$\Pi(\rho) = \sum_k p_k \rho_k = \sum_k P_k \rho P_k. \tag{8}$$

For example, if a projective measurement described by projectors $P_k$ is performed on a quantum system, what can we say about the Tsallis entropy of the system if we are not in position to see results of the measurement? If the state of the system before the measurement was $\rho$, then the state after the measurement is given by

$$\rho' = \sum_k P_k \rho P_k. \tag{9}$$

The following theorem states that the Tsallis entropy is never decreased by this procedure, and remains constant only if the state is not changed by the measurement.

*Theorem 1* (Projective measurements will not decrease the Tsallis entropy): Suppose $P_k$ is a complete set of orthogonal projectors and $\rho$ is a density operator. Then the entropy of the state $\rho' = \sum_k P_k \rho P_k$ of the system after measurement is at least as great as the original Tsallis entropy,

$$S_q(\rho') \geq S_q(\rho), \tag{10}$$

with the equality if and only if $\rho = \rho'$, where $S_q(\rho)$ represent the quantum Tsallis entropy of degree $q$ and it is defined as in (3)

$$S_q(\rho) = (1-q)^{-1} \left( \mathrm{tr}(\rho^q) - 1 \right), \quad q > 0, q \neq 1.$$

Proof:

Spectral decomposition of state $\rho$ can be expressed as

$$\rho = \sum_a p(a)|a\rangle\langle a| \tag{11}$$

where $p(a) \in [0, 1]$, $\sum_a p(a) = 1$, and $\{|a\rangle\}$ represents a complete orthonormal base. Based on this, we can express state $\rho'$ as

$$\rho' = \Pi(\rho) = \sum_k P_k \left(\sum_a p(a)|a\rangle\langle a|\right) P_k. \tag{12}$$

Writing $P_k = |k\rangle\langle k|$ with $\langle k|k\rangle = \delta_{kk'}$, we have ($\{|k\rangle\}$ represents a complete orthonormal base)

$$\Pi(|a\rangle\langle a|) = \sum_k |\langle a|k\rangle|^2 |k\rangle\langle k|. \tag{13}$$

Thus, we find

$$[\Pi(\rho)]^q = (\rho')^q = \sum_k \left[\sum_a p(a)|\langle a|k\rangle|^2\right]^q |k\rangle\langle k|, \tag{14}$$

which can lead to

$$\mathrm{tr}\left[(\rho')^q\right] = \sum_k \left[\sum_a p(a)|\langle a|k\rangle|^2\right]^q. \tag{15}$$

Here, we can note[5] that

$$\sum_a |\langle a|k\rangle|^2 = \sum_k |\langle a|k\rangle|^2 = 1. \tag{16}$$

The following analyzes will be split in two cases: $q > 1$ and $0 < q < 1$.

First lets analyze the case $0<q<1$. Note that the function, $f(x) = x^q$ ($x > 0$, $0 < q < 1$) is concave. So,

$$f\left(\sum_i \lambda_i a_i\right) \geq \sum_i \lambda_i f(a_i), \tag{17}$$

where $\lambda_i \in (0,1)$ and $\sum_i \lambda_i =1$. From this, (15) and (16), it follows

$$\mathrm{tr}\left[(\rho')^q\right] = \sum_k \left[\sum_a p(a)|\langle a|k\rangle|^2\right]^q \geq \sum_a p(a)^q \sum_k |\langle a|k\rangle|^2 = \sum_a p(a)^q. \tag{18}$$

This leads, after simple reasoning, to

$$S_q(\rho') \geq S_q(\rho), \tag{19}$$

which means, that projective measurement can only increase the Tsallis entropy of the state if $0 < q < 1$.

Now, we will analyze the case $q > 1$. In that case we can note that function $f(x) = x^q$ ($x > 0$, $q > 1$) is convex. So,

$$f\left(\sum_i \lambda_i a_i\right) \leq \sum_i \lambda_i f(a_i), \tag{20}$$

where $\lambda_i \in (0,1)$ and $\sum_i \lambda_i =1$. From this, (15) and (16), it follows

$$\mathrm{tr}\left[(\rho')^q\right] = \sum_k \left[\sum_a p(a)|\langle a|k\rangle|^2\right]^q \leq \sum_a p(a)^q \sum_k |\langle a|k\rangle|^2 = \sum_a p(a)^q. \tag{21}$$

Again, this leads, after simple reasoning, to

$$S_q(\rho') \geq S_q(\rho), \tag{22}$$

which means, that projective measurement can only increase the Tsallis entropy of the state if $q>1$.

So, now we can have a general conclusion that projective measurement can only increase (and cannot decrease) the Tsallis entropy of the state. It completes the proof.

This result can be extended to quantum unified $(r, s)$-entropies, using the same line of reasoning. In that case we can consider the following cases:

1. $0 < r < 1$ and $s > 0$,
2. $0 < r < 1$ and $s < 0$,
3. $r > 1$, and $s > 0$,
4. $r > 1$, and $s < 0$,
5. $r \neq 1$, and $s = 0$.

Using (18) and (21) and simple reasoning, it can be shown that projective measurement will not decrease the quantum unified ($r$, $s$)-entropy of the state.